\documentclass[conference]{IEEEtran}
\IEEEoverridecommandlockouts
\usepackage{cite}
\usepackage{amsmath,amssymb,amsfonts}
\usepackage{algorithmic}
\usepackage{graphicx}
\usepackage{textcomp}
\usepackage{xcolor}
\usepackage{booktabs}
\usepackage{multirow}
\usepackage{caption} % For controlling captions
\usepackage{subcaption}
\usepackage{makecell}
\usepackage{url}
\usepackage{lipsum}
\usepackage{booktabs} % For professional-looking tables
\usepackage{graphicx} % For adding images if needed
\usepackage[table,xcdraw]{xcolor} % For coloring cells
\usepackage{booktabs, multirow}
\usepackage[font=small]{caption}
\usepackage{adjustbox}
\usepackage{enumitem}

\def\BibTeX{{\rm B\kern-.05em{\sc i\kern-.025em b}\kern-.08em
    T\kern-.1667em\lower.7ex\hbox{E}\kern-.125emX}}
\begin{document}

\setlength{\belowcaptionskip}{-5pt}
\newcommand{\ms}[1]{{\color{orange}#1}}
\newcommand{\at}[1]{{\color{red}#1}}
% \title{Automating RTL generation using Agentic LLMs\\
\title{PEFA-AI: Advancing Open-source LLMs for RTL generation using \underline{P}rogressive \underline{E}rror \underline{F}eedback \underline{A}gentic-AI\\
% {\footnotesize \textsuperscript{*}Note: Sub-titles are not captured in Xplore and
% should not be used}
% \thanks{Identify applicable funding agency here. If none, delete this.}
}

\author{
\IEEEauthorblockN{}
\and
\IEEEauthorblockN{Athma Narayanan}
\and
\IEEEauthorblockN{Mahesh Subedar}
\\
\textit{Intel Corporation }\\
\textit{USA}
\and
\IEEEauthorblockN{Omesh Tickoo}
\and
\IEEEauthorblockN{}
% \IEEEauthorblockA{\textit{} \\
% \textit{}\\
% }
% }
}

\maketitle

\begin{abstract}
We present an agentic flow consisting of multiple agents that combine specialized LLMs and hardware simulation tools to collaboratively complete the complex task of Register-Transfer Level (RTL) generation without human intervention. A key feature of the proposed flow is the progressive error feedback system of agents (PEFA), a self-correcting mechanism that leverages iterative error feedback to progressively increase the complexity of the approach. The generated RTL includes checks for compilation, functional correctness, and synthesizable constructs. To validate this adaptive approach to code generation, benchmarking is performed using two open-source natural language-to-RTL datasets. We demonstrate the benefits of the proposed approach implemented on an open source agentic framework, using both open- and closed-source LLMs, effectively bridging the performance gap between them. Compared to previously published methods, our approach sets a new benchmark, providing state-of-the-art pass rates while being efficient in token counts.\footnote{The work appeared in the Design Automation Conference (DAC) 2025, Workshop Poster on June 22, 2025.}

\end{abstract}
% ##########################INTRO START
\section{Introduction}

%Generative AI has seen remarkable advancements in recent years, with large language models (LLMs) leading a revolution in the AI domain. Despite their transformative potential, LLMs have yet to completely eliminate the need for human involvement in real-world applications. One such domain is hardware design, a highly complex and multifaceted process that spans from architecture definition to chip tape-out. 
% This intricate workflow stands to benefit significantly from the adoption of an agentic framework.
% In this work, we explore the possibility of using agentic LLMs for RTL code generation.

% The complexity of hardware design is increasing exponentially with silicon technology scaling, while productivity remains constant~\cite{olofsson2018silicon}, resulting in a doubling of the cost for every generation of the process node. These tasks need human feedback and iteration, resulting in long development cycles. The development time for a typical system-on-chip product from definition to production takes around five to six quarters. With the doubling of transistors for every generation of the product while keeping the same time-to-market, exponentially more human effort is needed. This exponential growth in development cost is not sustainable. Hence, there is a push to incorporate advancements in AI technologies in hardware development process. 

Hardware design complexity is growing exponentially with silicon scaling, while productivity remains constant~\cite{olofsson2018silicon}, doubling costs per process node. These tasks require human feedback and iteration, leading to long development cycles - typically five to six quarters for a system-on-chip. Maintaining time-to-market while doubling transistors per generation demands exponentially more effort, making costs unsustainable. This drives the push to integrate AI advancements into hardware development.

In RTL code generation~\cite{lu2024rtllm}, users prompt LLMs with natural language and manually check for errors by compiling the code. Human-in-the-loop systems require iterative prompt refinement and corrections~\cite{thakur2023autochip} to guide the LLM toward the desired output. In self-correcting LLMs, compilation errors are iteratively fed back until convergence, but the most useful insights from error logs remain unclear. Large logs often overwhelm the model, leading to hallucinations~\cite{li2024dawn,simhi2025trust} and divergence. An automated system can enhance productivity by filtering relevant information and involving humans only when repeated trials fail to converge.

LLM driven agentic framework is well-suited for a variety of tasks, including RTL generation, verification, and design synthesis~\cite{zhao2024mage,ho2024verilogcoder}. By enabling agents to collaborate and interact, these tasks can be executed efficiently in a scalable manner. Agentic systems leverage LLM agents that collaborate with other agents, such as simulators, code executors, data indexers and so on. This collaborative approach not only enhances efficiency but also enables more robust and scalable solutions.

AutoGen~\cite{wu2023autogen}, LangChain~\cite{langchain2022}, and AgentVerse~\cite{agentverse2023} are among the open-source programming frameworks for agentic AI, designed to address these needs. AutoGen enables the creation and management of autonomous AI agents that can execute complex tasks, make decisions, and adapt to dynamic inputs. These frameworks provide the architecture and tools necessary for developing AI systems that operate independently while enabling effective collaboration among agents. LLMs~\cite{kapoor2024ai} excel at synthesizing information and defining the next steps for various agents. In the context of hardware (HW) design, AI agents are expected to interpret design requirements, generate RTL code, simulate performance, and suggest optimizations to meet target metrics with minimal human intervention. While agentic approaches have garnered significant attention, their efficiency and token consumption for RTL generation remain unbenchmarked, making it challenging to effectively compare different solutions.

Hence we propose a solution to address these challenges. Our contributions are as follows:
\begin{itemize}
% \item We propose an \ms{open-source} agentic framework built on  \textit{AutoGen}\cite{wu2023autogen} backbone that introduces syntactical, functional and test-pass aware error feedback to LLMs for self-correction. The system utilizes a  modular family of agents and can be extended to optimize the target power-performance-area (PPA)
% constraints and other RTL generation tasks such as code completion.
% \item We conduct a feasibility study and benchmark against two datasets (\textit{VerilogEval} \cite{liu2023verilogeval} and \textit{RTLLM1.1} \cite{lu2024rtllm}) and two categories of LLM models: closed-source models (\textit{GPT-4o} \cite{openai2023gpt4} and \textit{Claude-3.5-sonnet} \cite{}) and open-source instruct models (\textit{Llama 3.1:70B} and \textit{DeepSeekCoder:33B} \cite{dubey2024llama}). We compare their test-pass success rates using the \texttt{Pass@20} metric for text-to-RTL generation.
% \item We demonstrate the benefits of such a system in reducing mismatches between the expected test results and generated outputs and reducing the number of LLM calls. We showcase a technique to minimize the token usage by leveraging black-box test benches and summarized error log outputs.
\item We present an agentic framework that employs specialized LLM agents collaborating with hardware simulation tools to achieve RTL generation without human intervention. The framework uses a novel self-correcting system that implements iterative refinement through progressive error feedback agents to check for compilation and functional correctness using synthesizable RTL constructs. 
\item We conduct a thorough study to benchmark against two datasets  (\textit{VerilogEval}~\cite{liu2023verilogeval} and \textit{RTLLM1.1}~\cite{lu2024rtllm}) and both closed-source (\textit{GPT-4o}~\cite{openai2023gpt4} and \textit{Claude-3.5-sonnet}) and open-source (\textit{Llama 3.1:70B}~\cite{dubey2024llama} and \textit{DeepSeekCoder:33B}~\cite{deepseek-coder}) models. We demonstrate that the PEFA-AI approach benefits all the models and helps to bridge the gap between closed-source and open-source models. Our results show state-of-the-art performance compared to various benchmarks from the literature. The token counts from PEFA-AI results confirm the efficiency of the approach.
\end{itemize}

% ##########################INTRO END

\begin{figure*}[htbp]
\centerline{\includegraphics[scale=0.30]{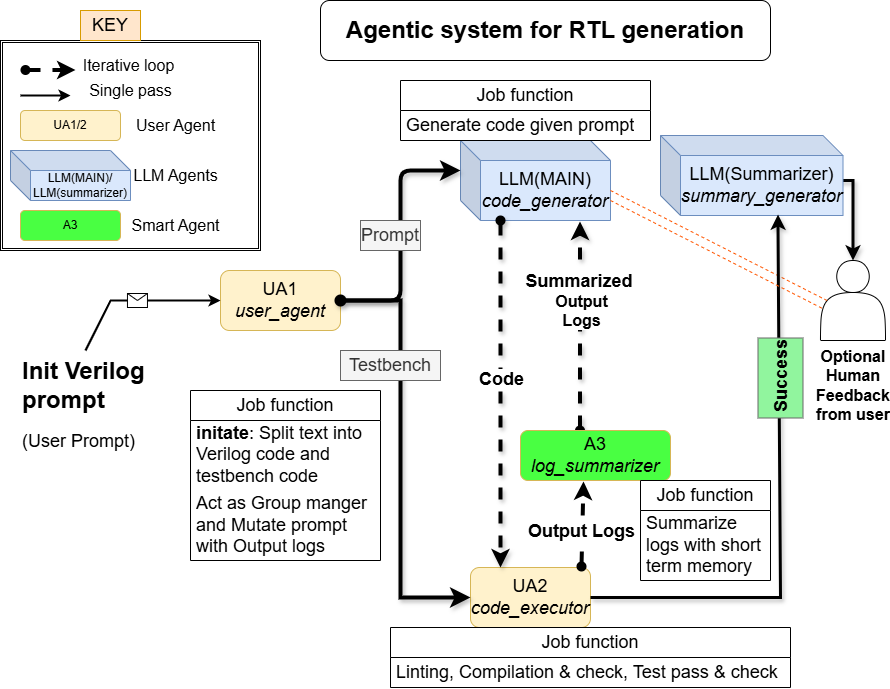}}
\caption{Overall Agentic workflow that takes an input prompt and test bench to generate the correct RTL code over $N$ agentic feedback loops.}
\label{fig:agentic_workflow}
\end{figure*}

% ##########################RELATED WORK START

\section{Related Work}
%Recent advancements in machine learning and natural language processing have enabled significant progress in areas such as code generation and test bench creation. Hardware design community is embracing the advancements in various forms such as hardware design space exploration, validation testing etc[NEEDS CITATION].\textcolor{red}{This para needs update and citations for good AI/LLM based techniques in Hardware}
Recent advancements in machine learning and natural language processing have enabled significant progress in many areas including hardware development~\cite{Arch2}.  The hardware design community is embracing these advancements in various forms including but not limited to hardware architecture \textit{design space exploration}~\cite{hong2023dosa}. These enable exploration of larger design spaces and potentially uncover more optimal solutions than traditional methods. Another area of interest is \textit{design verification}~\cite{dinu2019opportunities} which improves coverage-driven verification, test case generation, and debug processes. Furthermore, AI algorithms are being utilized to enhance physical design automation, timing closure, and design for manufacturability. AI-powered Electronic Design Automation (EDA) commercial tools~\cite{SynopsysRTL} are being developed for structural design~\cite{mirhoseini2020chip} to optimize chip designs for power-performance-area (PPA) constraints. 

\textbf{RTL generation using LLMs:}
Code generation using LLMs~\cite{jiang2024survey} encompasses several sub-problems, including code completion~\cite{li2022competition}, natural language-to-code translation~\cite{nijkamp2022codegen}, bug detection~\cite{kang2023large}, and identifying security vulnerabilities~\cite{yao2024survey}. However, the application of LLMs to RTL generation~\cite{thakur2024verigen,lu2024rtllm} has emerged only recently due to challenges such as the lack of high-quality datasets, scarcity of publicly available IP, and the absence of standardized benchmarks. Despite these hurdles, progress has been made in areas like PPA-aware RTL generation~\cite{thorat2023advanced,delorenzo2024make} and zero-shot code synthesis~\cite{sandal2024zero}. Additionally, metrics beyond functional correctness, such as creativity~\cite{delorenzo2024creativEval}, are beginning to be explored in this domain.

Monte Carlo Tree Search (MCTS)-based RTL generation~\cite{delorenzo2024make,zhang2023planning}, optimizes code generation by leveraging reward-driven search strategies to minimize the search space. This approach eliminates the need for a \texttt{pass@100} code generation paradigm by focusing on efficient exploration and exploitation of candidate solutions. Our implementation is shown in  Appendix Section \ref{sec:mcts}. However, these methods fall short in benchmarking agentic workflows, where feedback is provided more directly through natural language error logs. In this work, we evaluate the feasibility of self-correcting LLMs, which we argue are critical for enabling autonomous and robust RTL generation pipelines.

\textbf{Agentic Systems:}
% ##########################RELATED WORK END
Agents have proven to be formidable tools for code generation~\cite{ishibashi2024self,dong2024self} automating processes and minimizing the need for human intervention in the loop. The MAGE~\cite{zhao2024mage} study introduces a multi-agent framework designed to generate RTL. This framework employs varying temperatures and an RTL simulation tool to verify correctness. However, the results presented in the paper are unclear whether the reported pass rates pertain to the generated tests or the ground truth testbenches for the selected dataset. Additionally, the paper does not specify the number of steps the framework takes, which makes it difficult to compare this solution. To the best of our knowledge, using open-source tools, we are the first to investigate the efficacy of error feedback specifically for RTL generation.  This raises critical questions, such as which types of errors should be fed back and which LLMs are most suitable for this use case. To address these challenges, we benchmark this setup using our custom-designed agents.

\section{Proposed Framework}\label{Proposed framewoerk}

In conventional LLM-aided code generation systems, the process begins with the user providing a descriptive prompt to the LLM. The model generates multiple code candidates, typically using a \texttt{$\text{pass@}N$} (where $N \in [20, 100]$) strategy, where the best result—based on the highest average softmax score—is returned to the user. Once the code is received, the user compiles it alongside a test bench, and the process concludes. If the generated code fails to meet expectations, the user must manually iterate by either editing the original prompt to clarify their requirements or explicitly pointing out errors in the generated output. This iterative feedback loop relies heavily on human intervention to refine prompts and guide the system toward producing the desired results. To automate this process intelligently, an agentic system is required that takes both the prompt and the test bench as input. 

\begin{figure}
    \centering
    \includegraphics[width=0.45\textwidth]{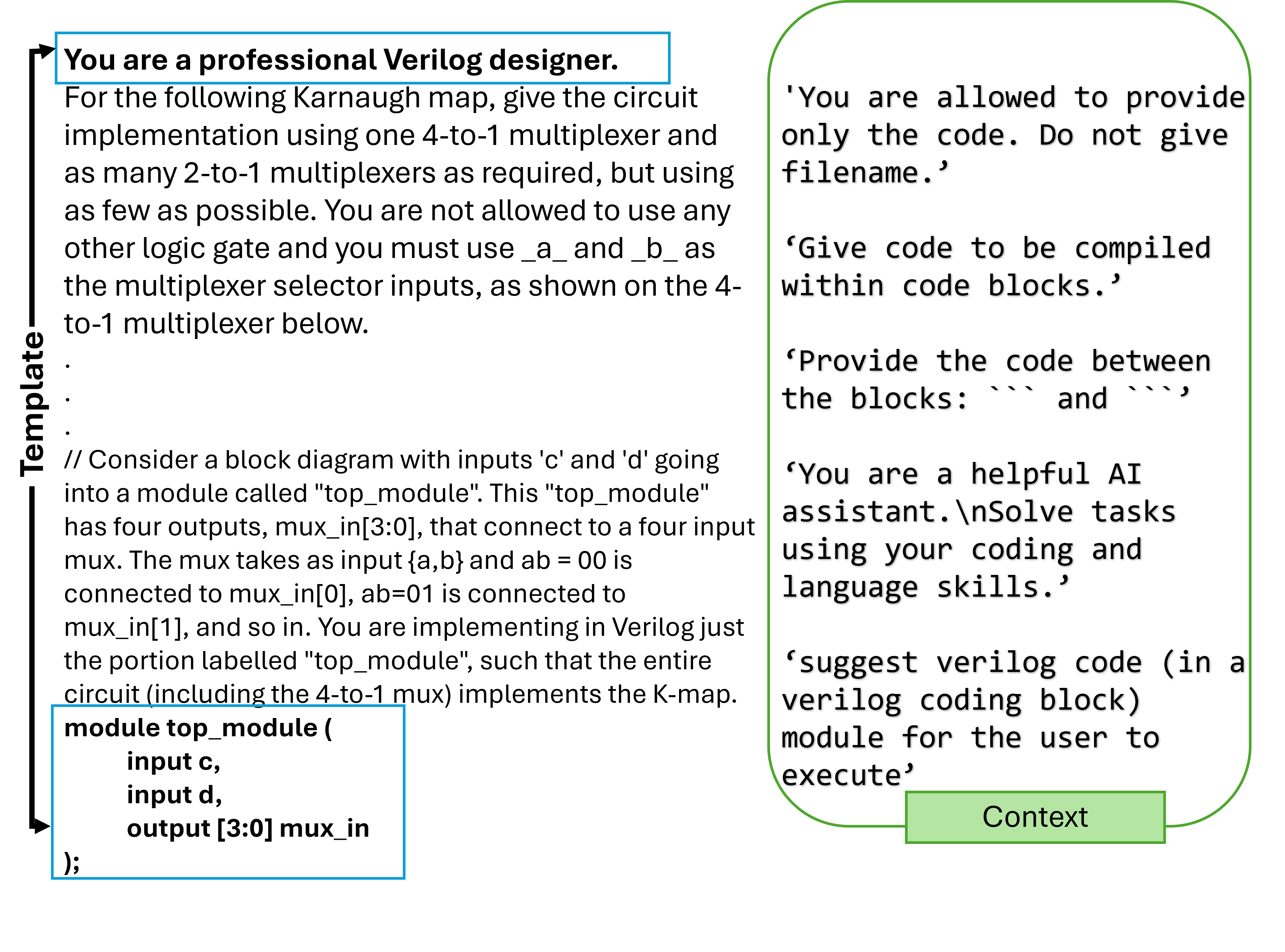}
    \caption{Illustrated here are the prompt and template setups. A context or system message is provided to the LLM, directing it to produce Verilog code, which is subsequently executed by the code executor agent.}
    \label{fig:templateprompt}
\end{figure}

The proposed agentic system for RTL generation can integrate with both open- and closed-sourced LLMs to streamline the hardware design process and privacy concerns of designers. AutoGen provides a robust framework designed to simplify the management of agents and enable efficient interaction with LLMs via API calls, abstracting and automating the underlying complexities. The framework's \texttt{Group\_chat\_manager} module facilitates collaborative workflows among multiple agents. In this study, we extend AutoGen by incorporating custom-built agents tailored to our specific requirements. The resulting system integrates multiple agents working in coordination, as illustrated in Figure~\ref{fig:agentic_workflow}. The system is built using a state-oriented approach, where the output of one agent triggers the appropriate response from the next agent in the sequence. This system iteratively refines the generated code by analyzing test results and intelligently updating the input prompt or logic until convergence or tests pass success. Precise templates used for prompt is showcased in  Figure~\ref{fig:templateprompt}. We describe the functions of each agent in detail and overall architecture below:

% \textcolor{green}{Take up 2 pages and explain fully.}
% \textcolor{red}{2 papras on why or how usual llms work in RTL and codegen. Show why our method diffeent}
% \textcolor{red}{Create subsections explaining each and every agent and reference the figures.}
% \textcolor{red}{Create an algorithm section to explaint he layout about poppoing and keeping the memory and hallucination if u keep all the context.)}

\subsection{Agent inputs}
In our design, the user inputs the prompt and test bench directly to the \textit{master\_agent}. The \textit{master\_agent} is programmatically tasked with parsing the user prompt and test bench. The parsed prompt is then directed to the \textit{code\_generator} agent. The \textit{code\_generator} agent leverages the LLM in the backend in a zero-shot manner~\cite{sandal2024zero,wang2023codet5plus}, capitalizing on its impressive zero-shot performance to convert prompts into code. 
The test bench, however, is programmatically modified to include the \$monitor command to print input/output signal changes along with the VCD dump file using a template.
%The test bench, however, is programmatically modified to include \textcolor{red}{monitor commands(write this better as it is important,maybe example in figure?)} in the output VCD using a template. 
This modification allows the system to generate an entire stack trace of I/O signal changes, significantly improving visibility into the generated code's behavior. This approach is notably more informative than the original test bench outputs in the VerilogEval dataset, which only report mismatches or indicate compilation success. Once modified, the test bench is communicated to the \textit{code\_executor} agent, which stores the test bench to a file for future compilation. \textit{code\_executor} agent, plays a critical role in parsing the outputs generated by the LLM. It is built on top of the \textit{LocalCommandLineCodeExecutor}  class from AutoGen and identifies specific keywords such as \textit{module} and \textit{endmodule} to structure the code into programmable RTL format. Additionally, the agent extracts code chunks enclosed in strings starting with triple quotes (\texttt{'''}) and appends them as necessary to ensure the generated code is complete and properly formatted.

% \begin{figure*}[htbp]
%     \centering
%     \begin{subfigure}[t]{0.45\textwidth}
%         \centering
%         \includegraphics[width=\linewidth]{pics/agent workflow2.png}
%         \caption{\raggedright Overall Agentic workflow that takes an input prompt and test bench to generate the correct RTL code over $N$ agentic feedback loops.}
%         \label{fig:agentic_workflow}
%     \end{subfigure}
%     \begin{subfigure}[t]{0.44\textwidth}
%         \centering
%         \includegraphics[width=\linewidth]{pics/Logsummarizer2.png}
%         \caption{\raggedright State flow diagram illustrating the layout of the log summarizer. The system allows a maximum of four feedback loops. Once the compilation succeeds, the counter $N$ starts. The feedback process begins by sending a basic error message to the LLM, specifically addressing mismatches in the VCD dump. If $N > 1$, an additional summarized feedback message is generated by the smaller LLM.}
%         \label{fig:Logsummarizer2}
%     \end{subfigure}
%     \caption{\raggedright Comparison of the agentic workflow and log summarizer state flow.}
% \end{figure*}

\subsection{Hybrid agent}
One of innovations we propose is to avoid exposing the test bench to the LLM, as we found no significant improvements from doing so, while it unnecessarily increases token counts. Instead, we treat the test bench as a black-box function, exposing only the output logs to open or closed-source models, thus preserving intellectual property (IP). This ensures efficiency and security without compromising performance.

We achieve this functionality by combining several smaller programmatic smart agents (a mixture of tools and LLMs) that we call as hybrid agents. As a first step, this involves,  utilizing linting commands in \textit{Verilator} with the \texttt{-Wall} flag to ensure syntactical correctness.  We employ the linting tool for static analysis of generated RTL to identify potential issues such as syntax errors, semantic problems, and non-synthesizable constructs. The code is then compiled using \textit{Icarus Verilog (iverilog)}, and the compiled module is passed to the test bench for execution. At any stage—linting, compilation, or testing—if a failure occurs, the process is immediately aborted, and the corresponding stack trace is collected.

\begin{figure}[h]
    \centering
    \includegraphics[width=0.75\linewidth]{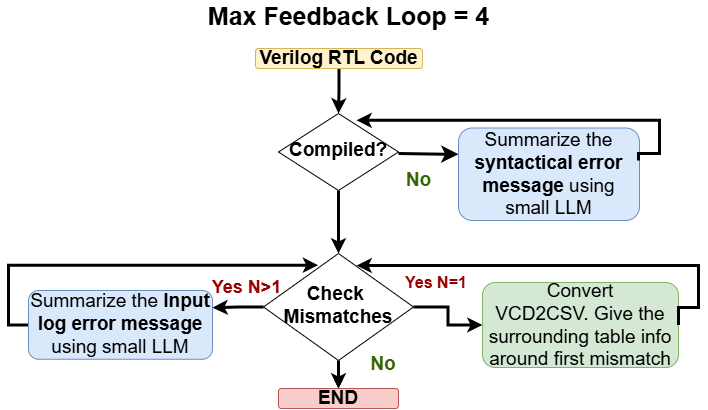} % Change filename accordingly
    \caption{
State flow diagram illustrating the layout of the log summarizer. The system allows a maximum of four feedback loops. Once the compilation succeeds, the counter $N$ starts.  The feedback process begins by sending a basic error message to the LLM, specifically addressing mismatches in the VCD dump. If $N > 1$, an additional summarized feedback message is generated by the smaller LLM.  %Both the compilation and the mismatch checker must complete within the four feedback loops. 
    }
    \label{fig:Logsummarizer2}
\end{figure}

Instead of feeding the full error logs in any step, we break down the problem-solving into multiple progressive error feedback steps. The following logic flow is used in the \textit{Log\_summarizer} agent to make it a progressive error feedback agent. For a maximum of $n=4$ feedback loops, we run the following logic check as described below and shown in Figure~\ref{fig:Logsummarizer2}:

\begin{enumerate}
    \item The process begins by checking whether the Verilog code compiles successfully. If compilation fails, a small LLM generates a summarized syntactical error message based on the logs.
    
    \item If the code compiles successfully, the next step is to check whether the feedback loop number equals 1.
    
    \item If \texttt{Loop Number = 1}, the simulation output \textit{VCD} file is converted to CSV using a parser. Additionally, surrounding table information around the first mismatch is provided for debugging.
    
    \item If \texttt{Loop Number $>$ 1}, a small LLM summarizes the input log error message to aid in debugging.
\end{enumerate}

The small LLM used in this process is a \textit{LLama-8B} model. Verilog VCD files provide the signal change waveform for the inputs and outputs from the test bench. From the VCD plots, we can determine the exact location of the mismatch between reference and simulation signals. The log summarizer agent extracts signal mismatches from the VCD file, which are then fed back to the RTL code generator. In Figure~\ref{fig:examplefig2}, we show the results from the generated RTL code that has errors and the output after four rounds of the agentic loops, feeding in the error information to correct the results. This progressive feedback helps the model concentrate on specific errors rather than contributing to an excessive token count, which can lead to hallucinations. In the results Section \ref{sec:progfeedback}, we provide the benefit of such an approach as an ablation study.  

\subsection{Context Manipulation}
It is common in agentic systems, particularly in a \textit{group\_chat} setting, to parse entire message histories as context. However, we found that when long outputs serve as feedback, the LLM often misinterprets text and fails to follow instructions accurately. This insight is crucial for the success of agentic systems. We address this issue in two ways: 

\begin{enumerate}
    \item \textit{Log\_summarizer} agent: Only the most recent outputs from the \textit{code\_executor} are sent to the \textit{log\_summarizer}, allowing it to act as an independent agent without being obfuscated by large amounts of data. Additionally, we clear all previous messages from the \textit{group\_chat\_manager}'s stack, sending only the most recent message for processing.
    
    \item Secondly, we ensure that only the most recent code that is passing compilation check is provided to the \textit{code\_generator}. To achieve this, we enqueue the generated codes along with its associated logs, allowing them to be processed and removed from the queue as necessary. This means the agent has visibility only to the latest best incorrect code with the latest \textit{log\_summarizer} agent's output. We manage the queue at each agent call, popping irrelevant messages and filtering the content to streamline the context.
\end{enumerate}
Finally, the summary of the entire \textit{ group\_chat} is generated and sent to the user, indicating the success or failure of the approach, using the \textit{summary\_generator}. This agent operates similarly to the \textit{log\_summarizer}, but acts on the entire group chat messages. By implementing such a one-of-a-kind agentic system, we can provide explainability to the user. This allows for minimal intervention by the designer, as the summarized outputs of the entire chat message provide insights into the agent's progress. This ensures that the LLM adheres to its initial instructions and avoids hallucinations. This feasibility study highlights a promising direction for further exploration and development and serves as a benchmark for more complex hardware design processes.

\subsection{Modular Agents}
The proposed agentic flow in Figure~\ref{fig:agentic_workflow} is modular and can be extended to include additional agents that can optimize the generated RTL for target power-performance-area (PPA) constraints beyond functional implementation. We can use commercial~\cite{OasysRTL,SynopsysRTL} or open-source~\cite{yosys} tools to obtain the logs on the PPA metric. The \textit{log\_summarizer} agent can provide feedback to the \textit{code\_generator} agent to generate RTL with target PPA constraints. The Silicon Compiler tool~\cite{SilCompiler}, which provides a Python wrapper for RTL synthesis, can be included as an agent to meet RTL PPA metric in the agentic flow. This is reserved for future work.

\section{Experimental Results and Evaluation}\label{ER}
\setlength{\belowcaptionskip}{0pt}
\begin{figure*}[htbp]
    \centering
    \includegraphics[width=0.93\textwidth]{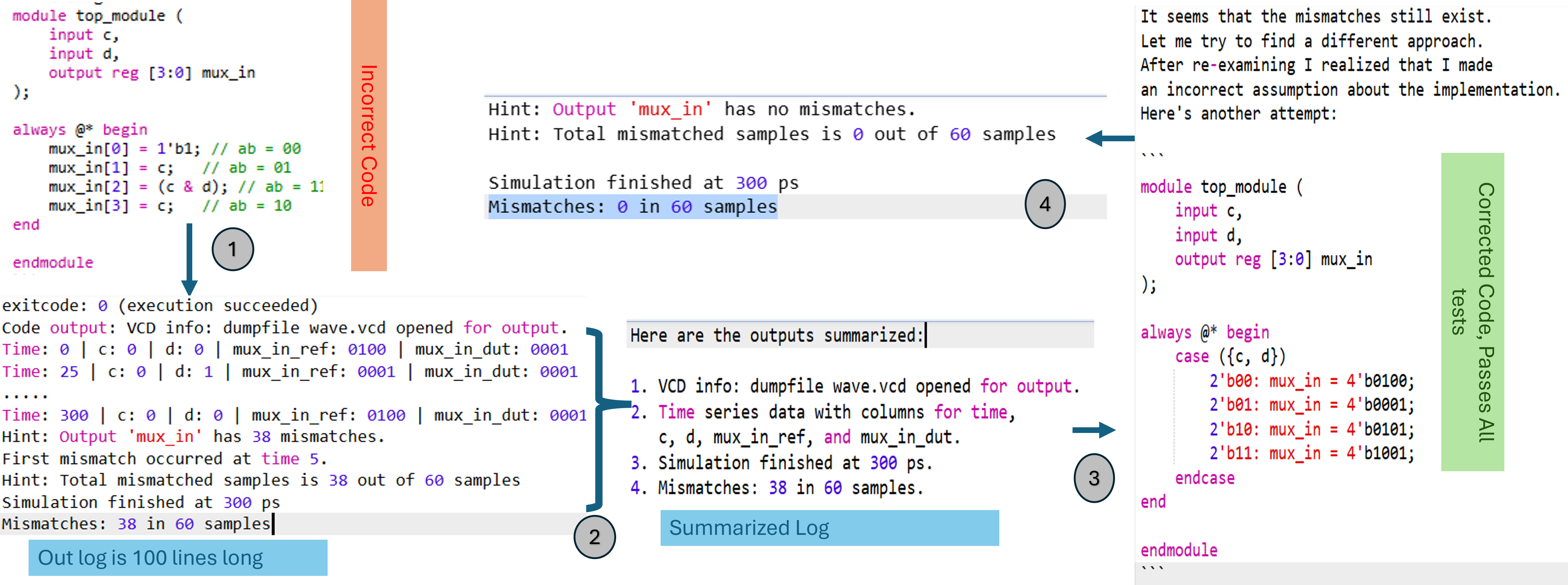}
    \caption{The incorrect code generated by the \textit{code\_{creator}} is executed by the \textit{code\_{executor}} agent. The test bench is modified using a template to generate all input and output values. As illustrated: (1) The output trace consists of hundreds of lines, requiring summarization. (2) The \textit{log\_{summarizer}} agent condenses the log into a few lines to prevent token explosion and reduce hallucination, aiding the LLM. This summary is fed back to the LLM in an iterative loop until successful compilation and test pass, as shown in (3). Subsequently, when the test passes successfully as in (4) with no mismatches, the run is completed.}
    \label{fig:examplefig2}
\end{figure*}
\setlength{\belowcaptionskip}{-5pt}

% \begin{figure*}[htbp]
% \centering
% \begin{subfigure}[t]{0.55\textwidth} % Slightly wider subfigure
%     \includegraphics[width=\textwidth]{pics/examplerun.png} % Match subfigure width
% \caption{Illustrated here are the prompt and template setups. A context or system message is provided to the LLM, directing it to produce Verilog code, which is subsequently executed by the code executor agent.}
%     \label{fig:sub1}
% \end{subfigure}
% \hspace{0.5\textwidth} % Reduce horizontal space
% \begin{subfigure}[t]{0.88\textwidth} % Slightly wider subfigure
%     \includegraphics[width=0.88\textwidth]{pics/examplerun2.png} % Match subfigure width
%     \caption{The incorrect code generated by the \textit{code\_{creator}} is executed by the \textit{code\_{executor}} agent. The test bench is modified using a template to generate all input and output values. As illustrated: (1) The output trace consists of hundreds of lines, requiring summarization. (2) The \textit{log\_{summarizer}} agent condenses the log into a few lines to prevent token explosion and reduce hallucination, aiding the LLM. This summary is fed back to the LLM in an iterative loop until successful compilation and test pass, as shown in (3). Subsequently when test pass is successful as in (4) with no mismatches the run is completed.}
%     \label{fig:sub2}
% \end{subfigure}
% \caption{Example of successful RTL generation using our agentic system \ms{can be split into two figures, top part can be one column}}
% \label{fig:example}
% \end{figure*}
\subsection{Implementation details}

\textbf{Datasets:} Natural language to RTL conversion is gaining significant attention due to its potential to streamline hardware design processes. Datasets such as VerilogEval~\cite{liu2023verilogeval,pinckney2024revisitingverilogevalnewerllms} and RTLLM1.1~\cite{lu2024rtllm} have been introduced as benchmarks for this purpose. These datasets are specifically designed to evaluate performance on complex prompts with higher token counts. In this study, we utilize these datasets, which contain 156 and 22 test set designs, respectively. Specifically we use both code completion and specification-to-rtl subsets of VerilogEval as our main benchmark. Following the setup in~\cite{ho2024verilogcoder}, we also remove approximately $7\%$ of erroneous test cases from the specification-to-rtl subset. RTLLM1.1, lacks a pre-existing test bench, which is critical for validating RTL designs. To overcome this challenge, we incorporate an expert human-in-the-loop approach to develop benchmarks. This ensures that at least one test is created for each test case, providing robust validation. This approach not only enables us to evaluate the successful completion of tasks but also allows us to measure improvements in rewards (or reductions in mismatches between the models' outputs and the expected ground truth). 
% \begin{table*}[!t]
% \caption{Results for VerilogEval and RTLLM1.1 Test Benches}
% \label{tab:results}
% \centering
% \begin{tabular}{lccccc}
% \toprule
% \bf{Dataset} & \bf{Method} & \bf{Model} & \bf{Passed/Failed} & \bf{Total} & \bf{LLM Calls} \\
% \midrule
% \multirow{4}{*}{\textbf{VerilogEval}} 
% & Pass@20 & Llama3-70B & 90/66 & 156 & 20 \\
% &  & ChatGPT & 104/52 & 156 & 20 \\
% \cmidrule(lr){2-6}
% & Ours & Llama3-70B & \textbf{98(+8)}/58 & 156 & 6 (Avg.) \\
% &  & ChatGPT & \textbf{106(+2)}/50 & 156 & 3 (Avg.) \\
% \midrule
% \multirow{4}{*}{\textbf{RTLLM1.1}} 
% & Pass@20 & Llama3-70B & 14/8 & 22 & 20 \\
% &  & ChatGPT & 18/4 & 22 & 20 \\
% \cmidrule(lr){2-6}
% & Ours & Llama3-70B & \textbf{17(+3)}/5 & 22 & 3 (Avg.) \\
% &  & ChatGPT & \textbf{19(+1)}/4 & 22 & 1 (Avg.) \\
% \bottomrule
% \end{tabular}
% \end{table*}
\begin{table*}[t]
    \centering
    \renewcommand{\arraystretch}{0.5} % Adjust row spacing
    \setlength{\tabcolsep}{4pt} % Adjust column spacing
    \rowcolors{3}{gray!15}{white} % Alternating row colors
\caption{Comparison of Pass rates (\%) for different models on Code Completion and Spec-to-RTL tasks using Non-Agentic and Agentic workflows. In the Non-Agentic workflow, the LLM is executed independently $n$ times without iterative refinement, while the Agentic workflow incorporates feedback loops to improve performance over multiple iterations.}
    \label{tab:model_comparison}
    
    % \begin{tabular}{lcccc|ccc||cccc|cccc}
    \begin{tabular}{lcccc|ccc!{\vrule\vrule\vrule}cccc|cccc}

        \toprule
        \multicolumn{1}{c}{} & 
        \multicolumn{7}{c}{\textbf{Code Completion}} \vrule\vrule\vrule &
        \multicolumn{7}{c}{\textbf{Spec-to-RTL}} \\
        % \multicolumn{7}{c|}{\textbf{Code Completion}} & \multicolumn{7}{c}{\textbf{Spec-to-RTL}}

      % \cmidrule{4}{c}
      \cmidrule(lr){2-9}
      \cmidrule(lr){9-15}
        \textbf{Model} & \multicolumn{4}{c}{Non-agentic} \vrule & \multicolumn{3}{c}{Agentic} \vrule\vrule\vrule  & \multicolumn{4}{c}{Non-Agentic} \vrule & \multicolumn{3}{c}{Agentic} \\
        \midrule

        \textit{LLM calls} & \textit{2} & \textit{3} & \textit{4} & \textit{20} & \textit{2} & \textit{3} & \textit{4} & \textit{2} & \textit{3} & \textit{4} & \textit{20} & \textit{2} & \textit{3} & \textit{4} \\
        \midrule
        GPT-4o & 66.66 & 68.58 & 69.2 & 75 & 76.3 & \textbf{76.92} & \textbf{76.92} & 68.7 & 68.7 & 70.23 & 77.09 & \textbf{77.83} & \textbf{77.83} & \textbf{77.83} \\
        GPT-4 & 55.12 & 59.61 & 60.9 & 74.3 & 78.8 & \textbf{80.13} & \textbf{80.13} & 67.93 & 70.92 & 70.92 & 77.86 & 79.38 & \textbf{80.91} & \textbf{80.91} \\
        Claude-3.5-sonnet & 65.38 & 69.87 & 72.4 & 78.2 & 80.1 & 83.33 & \textbf{84.61} & 74.8 & 77.86 & 78.62 & 81.67 & 87.02 & \textbf{90.07} & \textbf{90.07} \\
        Llama 3.1:8b Inst & 10.25 & 12.17 & 14.7 & 29.48 & 31 & 34 & \textbf{34.61} & 19.08 & 25.95 & 29.77 & 48.09 & 48.55 & 49.68 & \textbf{50.38} \\
        Llama 3.1:70b Inst & 48.71 & 55.12 & 58.3 & 68.58 & 70 & \textbf{72.19} & \textbf{72.19} & 55.725 & 60.3 & 62.59 & 71.75 & 74.8 & 74.8 & \textbf{75.5} \\
        Deepseekcoder:33b Inst & 35.25 & 39.74 & 44.9 & 57.69 & 60.90 & 60.90 & \textbf{62.17} & 32.82 & 37.4 & 40.45 & 59.54 & 61.82 & \textbf{63.35} & \textbf{63.35} \\
        \bottomrule
    \end{tabular}
    \label{table:results1}
\end{table*}

\textbf{Inference parameters:} Multiple open- and closed-source LLMs for evaluation for their prowess in code generation benchmarks are selected: open-source models, \textit{Llama3 70B} (Inst fine-tuned)~\cite{grattafiori2024llama},  \textit{DeepseekCoder:33B} (Inst fine-tuned), and closed-source model, \textit{ChatGPT-4o} and \textit{Claude-3.5-Sonnet~(2024-10-22)}. While many RTL code generation models exist~\cite{lu2024rtllm}, we are restricted to using only these models because of their instruction-following capabilities for self-correction. The agent interacts with all models through OpenAI API calls, enabling consistent query processing and comparison. To ensure uniformity in generation, we set the temperature parameter \(t = 0.8\) for all models and use \texttt{top\_k} sampling with \(k = 30\). These hyper-parameters were determined using a grid search for optimal performance. For the \textit{log\_summarizer} agent and the \textit{summary\_generator} agent, we use \textit{Llama3-8B}. These can be replaced by models with far lesser (sub-million) parameters~\cite{mishra2024attention} and do not add any significant compute as the primary purpose is summarization.

The overall \textit{group\_chat} functionality is configured to handle a maximum of 4 conversational feedback loops. This limit ensures controlled interaction flow and prevents excessive computational overhead as compared to pass@20. If a test pass is not achieved within these turns, the system automatically triggers the \textit{summary\_generator} agent. The summary consolidates the key outcomes of the interaction and informs the user of the success or failure in achieving the group task objectives. The human feedback can be optionally triggered in this case. Incorporating user feedback into the system’s iterative process could potentially enhance performance and adaptability. However, this remains beyond the scope of the current work and is left as a direction for future research.

\begin{table}[h]
    \centering
    \renewcommand{\arraystretch}{1.3} % Adjust row spacing
    \setlength{\tabcolsep}{6pt} % Increase column spacing for readability
    % \captionsetup{font=normal} % Increase caption font size

    \caption{Comparison of LLM performance based on pass rate (\%) with published baselines, using 20 LLM calls. For a fair comparison, we also run the agentic loops with $N = 20$. }

    \label{tab:llm_comparison}
    \begin{adjustbox}{max width=\columnwidth}
    {\huge % Change table text size
    \begin{tabular}{lccccc}
        \toprule
        \multirow{2}{*}{\textbf{Model}} & \multirow{2}{*}{\textbf{Agentic}} & \multicolumn{2}{c}{\textbf{VerilogEval}} & \multirow{2}{*}{\textbf{RTLLM1.1}} \\
        \cmidrule(lr){3-4}
        & & \textbf{Code complete} & \textbf{Spec-to-rtl} & \\
        \midrule
         {\huge Llama3.1:70B Inst} & {\huge N} & {\huge 68.58} & {\huge 71.75} & {\huge 63.00} \\
         {\huge DeepSeekCoder:33B Inst} & {\huge N} & {\huge 46.9} & {\huge 44.18} & {\huge 46.00} \\
         {\huge GPT-4o} & {\huge N} & {\huge 75.00}  & {\huge 77.10} & {\huge 71.81} \\
         {\huge Claude-3.5-sonnet} & {\huge N} & {\huge 78.21} & {\huge 81.68} & {\huge 77.27} \\
         % {\huge MAGE}~\cite{zhao2024mage}{\huge\textbf{\raisebox{1pt}{\textsuperscript{\huge [1]}}}} & {\huge Y} & {\huge \textbf{94.8}} & {\huge 95.7} & {\huge N/A} \\
         {\huge AIVRIL (Claude 3.5)}~\cite{sami2024aivril}{\huge\textbf{\raisebox{1pt}{\textsuperscript{\huge [1]}}}} & {\huge Y} & {\huge 70.51} & {\huge N/A} & {\huge N/A} \\
         {\huge OriGEN}~\cite{cui2024origen}{\huge\textbf{\raisebox{1pt}{\textsuperscript{\huge [1]}}}} & {\huge Y} & {\huge 54.4} & {\huge N/A} & {\huge 65.5} \\
         {\huge VerilogCoder}~\cite{ho2024verilogcoder}{\huge\textbf{\raisebox{1pt}{\textsuperscript{\huge [1]}}}} (Llama3) & {\huge Y} & {\huge N/A} & {\huge 67.3} & {\huge N/A} \\
         {\huge VerilogCoder}~\cite{ho2024verilogcoder}{\huge\textbf{\raisebox{1pt}{\textsuperscript{\huge [1]}}}} (GPT-4 Turbo) & {\huge Y} & {\huge N/A} &{\huge 94.8} & {\huge N/A} \\
\midrule
\midrule

{\huge Llama3.1:70B Inst Progressive}\textbf{(Ours)} & {\huge Y} & {\huge 79.49} & {\huge 78.6} & {\huge 77.27}\\
{\huge DeepSeekCoder:33B Inst Progressive}\textbf{(Ours)} & {\huge Y} & {\huge 64.1} & {\huge 63.35} & {\huge 51.27} \\
{\huge GPT-4o Progressive}\textbf{ (Ours)} & {\huge Y} & {\huge 83.97} & {\huge 81.68} & {\huge 86.83} \\
{\huge Claude-3.5-sonnet Progressive}\textbf{(Ours)} & {\huge Y} & {\huge {85.9}} & {\huge {90.83}} & {\huge {87.10}} \\

{\huge Llama3.1:70B Inst Progressive\textbf{(Ours)}{\huge\textbf{\raisebox{1pt}{\textsuperscript{\huge [1]}}}}}
 & {\huge Y} & {\huge 79.49} & {\huge 84.00} & {\huge 77.27} \\

{\huge DeepSeekCoder:33B Inst Progressive\textbf{(Ours)}{\huge\textbf{\raisebox{1pt}{\textsuperscript{\huge [1]}}}}} & {\huge Y} & {\huge 64.1} & {\huge 67.80} & {\huge 51.27} \\

{\huge GPT-4o Progressive\textbf{(Ours)}{\huge\textbf{\raisebox{1pt}{\textsuperscript{\huge [1]}}}}} & {\huge Y} & {\huge 83.97} & {\huge 86.98} & {\huge 86.83} \\

{\huge Claude-3.5-sonnet Progressive\textbf{(Ours)}{\huge\textbf{\raisebox{1pt}{\textsuperscript{\huge [1]}}}}} & {\huge Y} & {\huge \textbf{85.9}} & {\huge \textbf{97.95}} & {\huge \textbf{87.10}} \\
        
\bottomrule
    \end{tabular}
}

\end{adjustbox}

\vspace{7pt}
    \captionsetup{justification=justified, singlelinecheck=false}
\caption*{\footnotesize \textbf{Note:} \textbf{[1]} \textit{VerilogCoder\cite{ho2024verilogcoder} removes 7\% of erroneous specification-to-RTL problems from the 156 original problems. We keep the same setup and re-run the agentic workflows on the same dataset. For completeness we also show the performance on the full 156 problems as well.}
}
\label{tab:results2}
\end{table}
% \vspace{-3pt}

% \begin{figure}[h]
%     \centering
%     \includegraphics[width=0.75\linewidth]{pics/MCTS.png} % Change filename accordingly
%     \caption{
%     MCTS based approach to determine the optimal RTL code that passes all test cases.
%     %Both the compilation and the mismatch checker must complete within the four feedback loops. 
%     }
%     \label{fig:mcts}
% \end{figure}

\subsection{Results Discussion}
We present the results of our experiments conducted on the VerilogEval dataset, as summarized in Table \ref{table:results1}. In this evaluation, we analyze the number of LLM calls and compare their impact on performance. As discussed earlier, a single agentic cycle is constrained to a maximum of four LLM calls. Consequently, our baseline consists of a non-agentic approach, wherein there is no feedback from the execution trace. Instead, the prompt is executed through the model iteratively for $n=2$ to $n=4$ runs, ensuring a fair and consistent comparison. Our proposed solution demonstrates notable performance improvements across both tasks using both closed-source and open-source models. The agentic approach improves test pass rates, as shown in Table \ref{table:results1}, demonstrating enhanced problem-solving capabilities over non-agentic methods. Closed-source models such as GPT-4o, GPT-4, and Claude-3.5-Sonnet achieve test pass increases of approximately 11.11\%, 31.58\%, and 16.81\%, respectively. Meanwhile, open-source models like Llama 3.1:8B Inst, Llama 3.1:70B Inst, and DeepSeekCoder-33B attain increases of approximately 134.78\%, 24.18\%, and 38.57\%, respectively. This highlights the performance gains enabled by agentic reasoning, effectively bridging the gap between open-source and closed-source models. Furthermore, in the case of twenty LLM calls, where success is defined as at least one pass out of the twenty calls, we observe an improvement across all models. Importantly, our approach significantly reduces the number of calls made to the LLMs compared to the substantially higher number of calls required by \texttt{pass@20}. Table \ref{tab:results2} presents the \texttt{pass@20} results when performing twenty agentic runs. Our method achieves state-of-the-art results on the VerilogEval and RTLLM1.1 datasets. Compared to the results in Table \ref{table:results1}, running the agentic workflow twenty times shows minimal improvement compared to running it once, as shown in Table \ref{tab:results2}. From a computational efficiency standpoint, reducing API calls enhances our method's performance, making it a promising foundation for self-correcting LLMs. 
A successful run for creating a Karnaugh map is shown in Figure~\ref{fig:examplefig2} and associated VCD plot is shown in Appendix Section \ref{sec:appendixvcdplot}.

\subsection{Effect of Progressive feedback}
\label{sec:progfeedback}
\begin{figure}[htbp]
    \centering
    \includegraphics[width=0.7\linewidth]{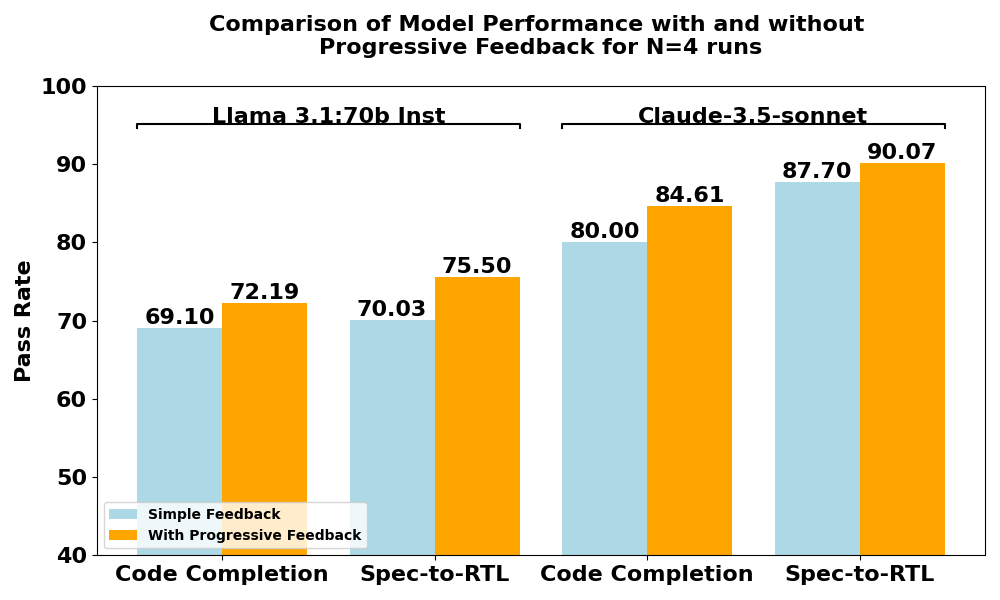}
    \caption{Bar plot showcasing the effect of progressive feedback on pass rate (\%). In the simple feedback setting the agent is only exposed to the surrounding mismatches from the \textit{log\_summarizer} four times.}
    \label{fig:feedbackcomp}
\end{figure}

% We compare the effect of progressive feedback, as shown in Figure \ref{fig:feedbackcomp}. Here, simple feedback refers to the scenario where the model is exposed only to the surrounding mismatch region in each feedback loop, without the use of \textit{log\_summarizer}.
% The effect of progressive feedback on model performance demonstrates a consistent improvement across tasks and models. For the \textit{Llama 3.1:70b Inst} model, the performance in the \textit{Code Completion} task improved by ${4.5\%}$, while in the \textit{Spec-to-RTL} task, the improvement was a ${7.8\%}$ gain. Similarly, the \textit{Claude-3.5-sonnet} model exhibited even greater improvements, with \textit{Code Completion} displaying a ${5.8\%}$ gain and \textit{Spec-to-RTL} rising from ${87.7}$ to ${90.7}$ (${2.7\%}$ increase). These results indicate that progressive feedback enhances model learning, particularly benefiting high-performing models. The more substantial improvements in the \textit{Spec-to-RTL} task for Llama 3.1:70b Inst suggest that feedback mechanisms are especially valuable for structured reasoning and complex transformations. This analysis underscores the role of iterative refinement in improving LLM-driven code generation and specification-based tasks.
We analyze the impact of progressive feedback, illustrated in Figure \ref{fig:feedbackcomp}. Simple feedback exposes the model only to the surrounding mismatch region without using \textit{log\_summarizer}. Progressive feedback consistently improves performance across tasks and models. For \textit{Llama 3.1:70b Inst}, \textit{Code Completion} improved by ${4.5\%}$, while \textit{Spec-to-RTL} saw a ${7.8\%}$ gain. \textit{Claude-3.5-sonnet} showed even greater improvements, with \textit{Code Completion} gaining ${5.8\%}$ and \textit{Spec-to-RTL} increasing from ${87.7}$ to ${90.7}$ (${2.7\%}$). These results highlight the effectiveness of iterative feedback, particularly for complex reasoning tasks like \textit{Spec-to-RTL}, reinforcing its role in enhancing LLM-driven code generation. More importantly unlike \cite{zhao2024mage,ho2024verilogcoder} that rely on heavily tuned temperature heuristics, our method is robust to temperature variation and provides consistent results as shown in Appendix Section \ref{sec:appendeixtemp}.

\subsection{Token count comparison}
We analyzed token usage in both the \texttt{pass@20} baseline and our agentic workflow to assess computational efficiency. For Claude-3.5-Sonnet, \texttt{pass@20} generates an average of $440$ total tokens per call~(with $390$ input tokens input $+$ $50$ output tokens). Our agentic approach, while maintaining similar token generation per problem, differs in input composition: the initial LLM call starts with $440$ tokens, followed by an average $600$ additional tokens per iteration from summarized logs of compiled RTL code. This results in an average of $1040$ total tokens per call, totaling $3,560$ tokens over four calls ($440~+~3\times 1040$). Despite the $600$ token overhead per call, our approach reduces LLM calls from $20$ to $4$, to find the passing candidate significantly improving efficiency. 
\begin{table}[h]
    \centering
\caption{Average total token count across problems in the VerilogEval dataset after four LLM calls (excluding calls to the small Log\_summarizer agent).}
    \begin{tabular}{lcc}
        \toprule
        Model & Not-Agentic & Agentic \\
        \midrule
        Llama 3.1: 70b Inst & 1760 &  3296\\
        Claude-3.5-sonnet & 1760 &  3560\\
        \bottomrule
    \end{tabular}

    \label{tab:token_counts}
\end{table}

While a direct token comparison may be unfair due to architectural differences—\texttt{pass@20} generates independent solutions, whereas our workflow refines them iteratively—our results highlight the advantages of agentic reasoning. Prioritizing functional correctness and memory efficiency, our method provides a flexible foundation for integrating additional time consuming agents, such as PPA-aware RTL synthesis, testbench generation, and RTL validation. 
Detailed per-model token comparisons are showcased in Appendix Section \ref{sec:appendixtokencounts}. 

\section{Conclusion and Future Work}
% In this work, we address the benefits of agentic flows for enabling self-correcting mechanisms in LLMs. Our study highlights a promising direction for leveraging such flows in RTL code generation tasks. Incorporating a progressive error correction approach can improve performance while reducing computational overhead. However,  fine-tuning \textit{log\_summarization} LLM agents to suggest possible corrections and mimic human intelligence remains an open challenge. Additionally, expanding the scope of benchmarking to include a broader range of  more complex RTL tasks such as RTL code completion, error handling and test bench creation is essential. These directions highlight the need for continued research to unlock the full potential of self-correcting LLMs in RTL code generation.
In this work, we address the benefits of agentic flows for enabling self-correcting mechanisms in LLMs. Our study highlights a promising direction for leveraging such flows in RTL code generation tasks. Incorporating a progressive error correction approach can improve performance while reducing computational overhead. However, fine-tuning \textit{log\_summarization} LLM agents to suggest possible corrections and mimic human intelligence remains an open challenge. Our method significantly closes the gap between open-source and closed-source solutions while achieving state of the art results. These directions highlight the need for continued research to unlock the full potential of self-correcting LLMs in RTL code generation. Additionally, expanding the scope of benchmarking to include a broader range of more complex RTL tasks, error handling, and test bench creation, is essential.

{\small
\bibliographystyle{ieeetr}
\bibliography{main}
}

\clearpage
\appendix
\section{Appendix}

\title{Appendix}

\subsection{Glossary}
\begin{enumerate}[label=\arabic*.]  % Forces standard numbering
    \item \textit{Pass@N}: A metric used in evaluating code generation models. It measures the probability that at least one of the top \( N \) generated solutions is correct. The formula is given by:

    \[
    \text{Pass@N} = 1 - \prod_{i=1}^{N} \left( 1 - p_i \right)
    \]

    where \( p_i \) represents the probability of the \( i \)-th generated solution being correct.

    Alternatively, using a sampling-based approximation:

    \[
    \text{Pass@N} = 1 - \frac{\binom{M - C}{N}}{\binom{M}{N}}
    \]

    where:
    \begin{itemize}
        \item \( M \) is the total number of generated samples,
        \item \( C \) is the number of correct solutions,
        \item \( \binom{M}{N} \) represents the binomial coefficient.
    \end{itemize}
   \item \textit{AutoGen}: AutoGen is an open-source framework developed by Microsoft that enables the creation of autonomous AI agents capable of collaborating and communicating to solve complex tasks. It is built on top of Large Language Models (LLMs) and is designed to facilitate multi-agent interactions, automation, and optimization in AI-driven workflows. 

We utilize the \textit{v0.2} agentic framework and modify the \textit{CommandLineExecutor} class to support Verilog linting and compilation. Additionally, we override the \textit{message\_parsing} method in the \textit{group\_chat} manager class to retain only the previously generated incorrect code while discarding other message history and logging summaries.

For more details, refer to the official repository:  
\url{https://github.com/microsoft/autogen}.

\item \textit{RTL}: Register Transfer Level
\item \textit{HW}: Hardware
\item \textit{LLM}: Large Language model.
\item \textit{PPA}: Power, Performance, and Area.
\item \textit{VCD}: Value Change Dump.
\end{enumerate}

\subsection{MCTS}
\label{sec:mcts}
In our baseline implementation, we adopt a Monte Carlo Tree Search (MCTS) framework to guide the seach for optimal RTL code that passes all test cases. Each node represents a partial program state, and its children correspond to possible next tokens predicted by the LLM. During the selection phase, the next child node is chosen using a PUCT-based strategy\cite{delorenzo2024make}, where the average reward term reflects the \emph{test pass rate} obtained from previous rollouts of that action. Specifically, actions that historically lead to higher test pass rates are favored, while the exploration term encourages visiting less explored actions based on their prior probabilities from the LLM. This allows the search to balance between exploiting high-performing code patterns and exploring new promising continuations.

\begin{figure}[h]
    \centering
    \includegraphics[width=0.8\linewidth]{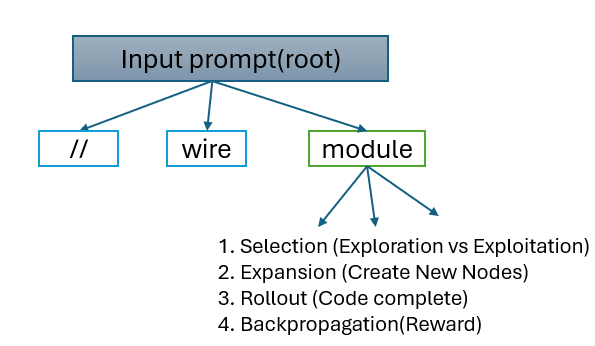}
    \caption{MCTS allows us to search for the optimal RTL code that passes all the test cases.}
    \label{fig:strbarplot}
\end{figure}

The reward function is defined based on the outcome of functional testing. Specifically, the reward is computed as the ratio between the number of test cases that pass and the total number of test cases. Formally, if $n_{\text{pass}}$ and $n_{\text{total}}$ denote the number of passed and total test cases respectively, the reward is given by
\[
R = 
\begin{cases}
\dfrac{n_{\text{pass}}}{n_{\text{total}}}, & \text{if } n_{\text{pass}} > 0, \\
-1, & \text{otherwise.}
\end{cases}
\]
This design penalizes code that fails all tests while proportionally rewarding partial correctness, thereby guiding the search toward progressively functional solutions.

\begin{table}[h!]
\centering
\caption[Performance Comparison]{Performance comparison on the VerilogEval dataset using MCTS and agentic flow. The first column shows the number of cases with improved rewards out of the total failed cases compared to \texttt{pass@1}. The second column presents the average LLM calls(or rollouts) needed.}
\begin{tabular}{|c|c|c|}
\hline
\textbf{Model} & \makecell{\textbf{Improved} \\ \textbf{/Failed}} & \makecell{\textbf{Avg. LLM} \\ \textbf{calls}} \\ \hline
Llama3-70B (MCTS) & 3/61 & 10 \\ \hline
Llama3-70B (Ours) & 8/58 & 4 \\ \hline
\end{tabular}

\label{table:improvement_data}
\end{table}

From the experimental results, it is evident that incorporating error feedback significantly improves the number of test cases passed compared to the standard MCTS approach. This demonstrates that iterative output refinement guided by feedback leads to more functionally correct and robust code generation.

\subsection{Mismatch rates}

\begin{table}[h!]
\centering
\caption[Performance Comparison]{Performance comparison on the VerilogEval dataset using both models and agentic flow. The first column shows the number of cases with improved rewards out of the total failed cases compared to \texttt{pass@1}. The second column presents the average reduction in mismatches achieved by each model.}
\begin{tabular}{|c|c|c|}
\hline
\textbf{Model} & \makecell{\textbf{Improved} \\ \textbf{/Failed}} & \makecell{\textbf{Avg. Reduction} \\ \textbf{in Mismatches}} \\ \hline
Llama3-70B (Ours) & 8/58 & 30 \\ \hline
Calude-3.5-Sonnet (Ours) & 15/50 & 582 \\ \hline
\end{tabular}

\label{table:improvement_data}
\end{table}

It is encouraging to observe that, even though some cases do not completely pass all the tests, our methodology significantly reduces mismatches (or increase in test pass rate) in the outputs of generated code in comparison to the test bench. This suggests that the generated code achieves higher rewards, indicating progress in the correct direction. To quantify this improvement, with respect to single shot code generation, we propose averaging the reduction in mismatches for each case with decreased mismatches. For the VerilogEval dataset, we present the averaged reduction of mismatches (or improved rewards) achieved using the agentic workflow, as shown in Table~\ref{table:improvement_data}. Higher reduction is better. Overall, while both models show benefits from the agentic workflow, Calude-3.5-Sonnet displays the biggest improvement, both in terms of the number of improved cases and the magnitude of mismatch reduction. Nevertheless, the data also reveals that both models leave many failed cases unresolved, emphasizing the need for further optimization and quality error feedback.

\subsection{VCD plot of Karnaugh map}
\label{sec:appendixvcdplot}
\begin{figure}[h]
\centering
\begin{subfigure}[t]{0.99\linewidth} % Slightly wider subfigure
  \includegraphics[width=\linewidth]{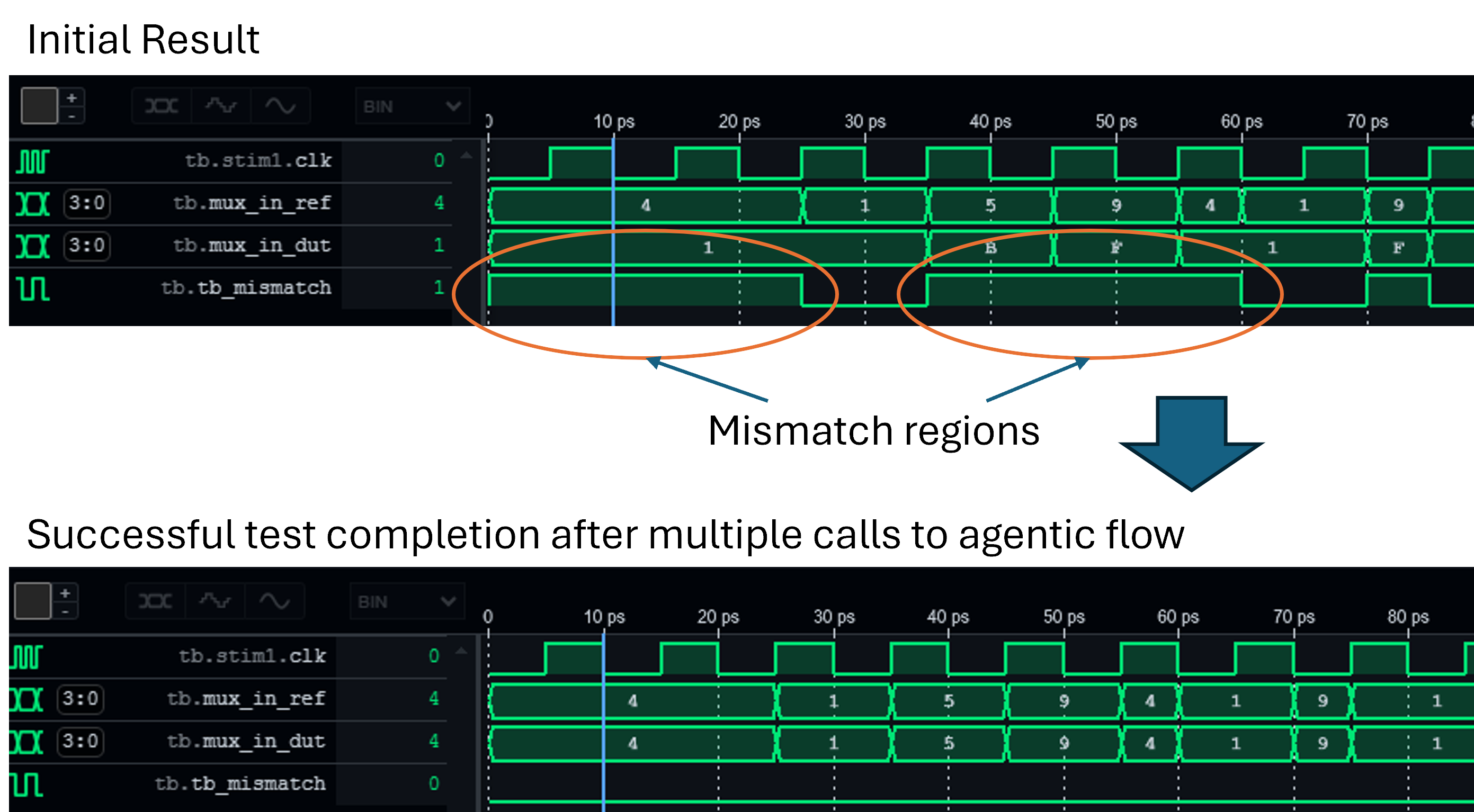} % Match subfigure width
\end{subfigure}
\caption{VCD plot (top) showcasing the mismatch regions for the generated RTL code. After three rounds of agentic flow, feeding in the VCD error information using the log summarizer, passes the test case (bottom).}
\label{fig:vcd_plot}
\end{figure}

We showcase an example VCD plots before and after the Agentic feedback loop in Figure \ref{fig:vcd_plot}.

\subsection{Effect of temperature on agentic runs}
\label{sec:appendeixtemp}
\begin{figure}[h]
    \centering
    \includegraphics[width=0.8\linewidth]{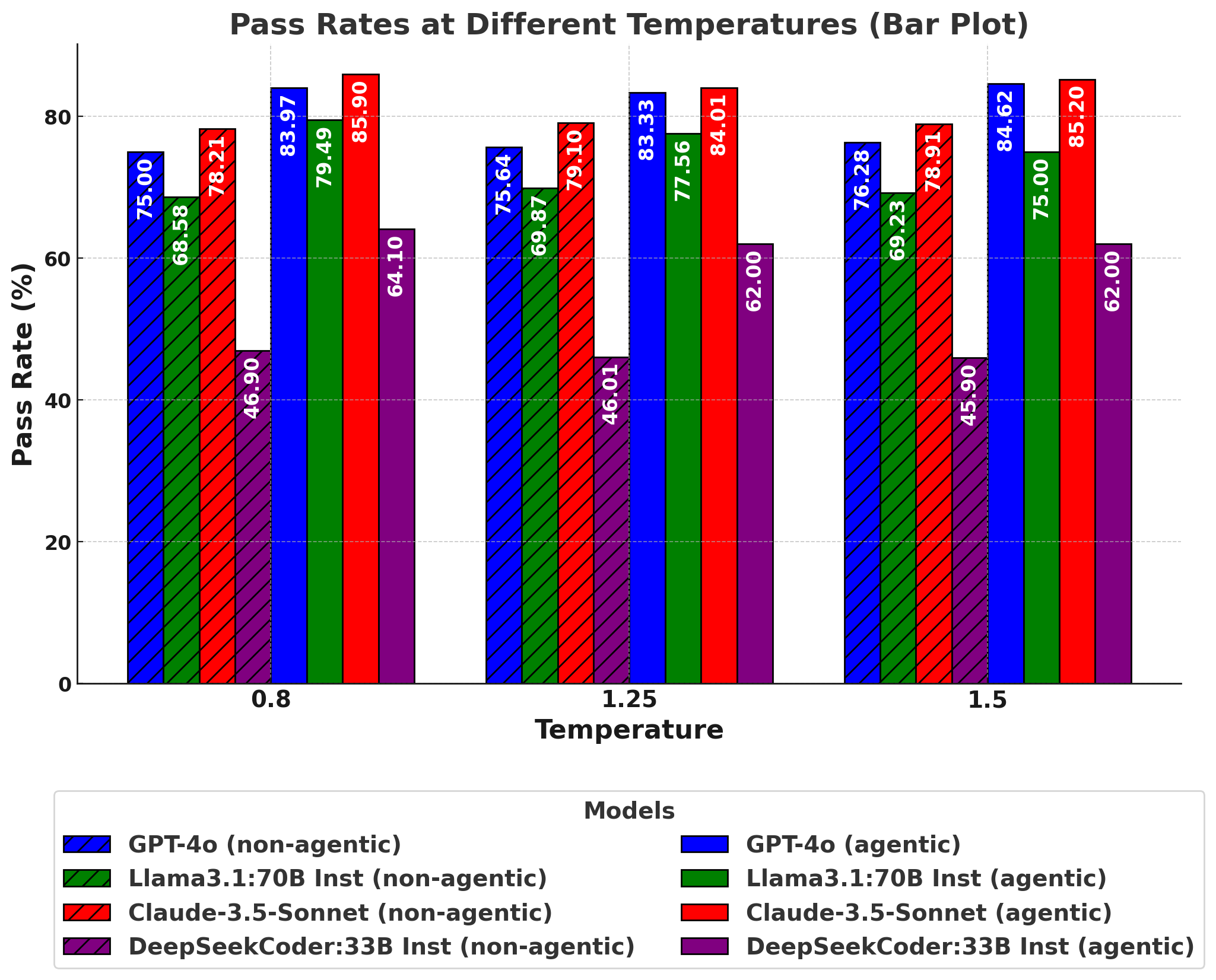}
    \caption{Pass rate over temperature on code-completion dataset}
    \label{fig:ccbarplot}
\end{figure}

\begin{figure}[h]
    \centering
    \includegraphics[width=0.8\linewidth]{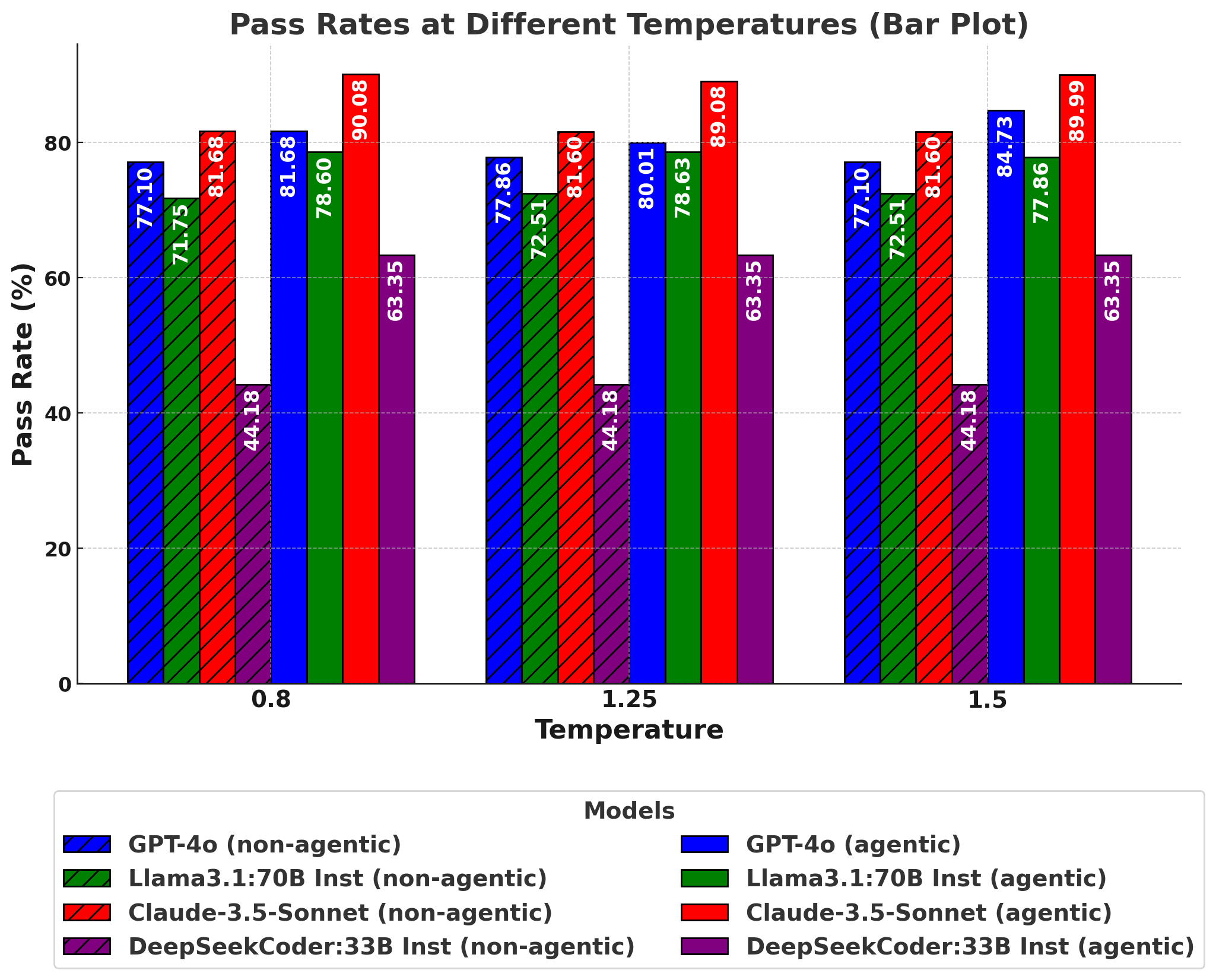}
    \caption{Pass rate over temperature on specification-to-rtl dataset}
    \label{fig:strbarplot}
\end{figure}

We conduct ablation study on the temperature value. Unlike \cite{zhao2024mage,ho2024verilogcoder} that require high temperature sampling we show that high temperature has little effect on the  final performance. However temperature=$0.8$ works best over all the models. \textit{Claude-3.5-Sonnet} API allows for values between $0$ and $1$. Hence we scale the temperatures greater than $1$ to support Anthropic API.
\begin{figure}[h]
    \centering
    \includegraphics[width=0.85\linewidth]{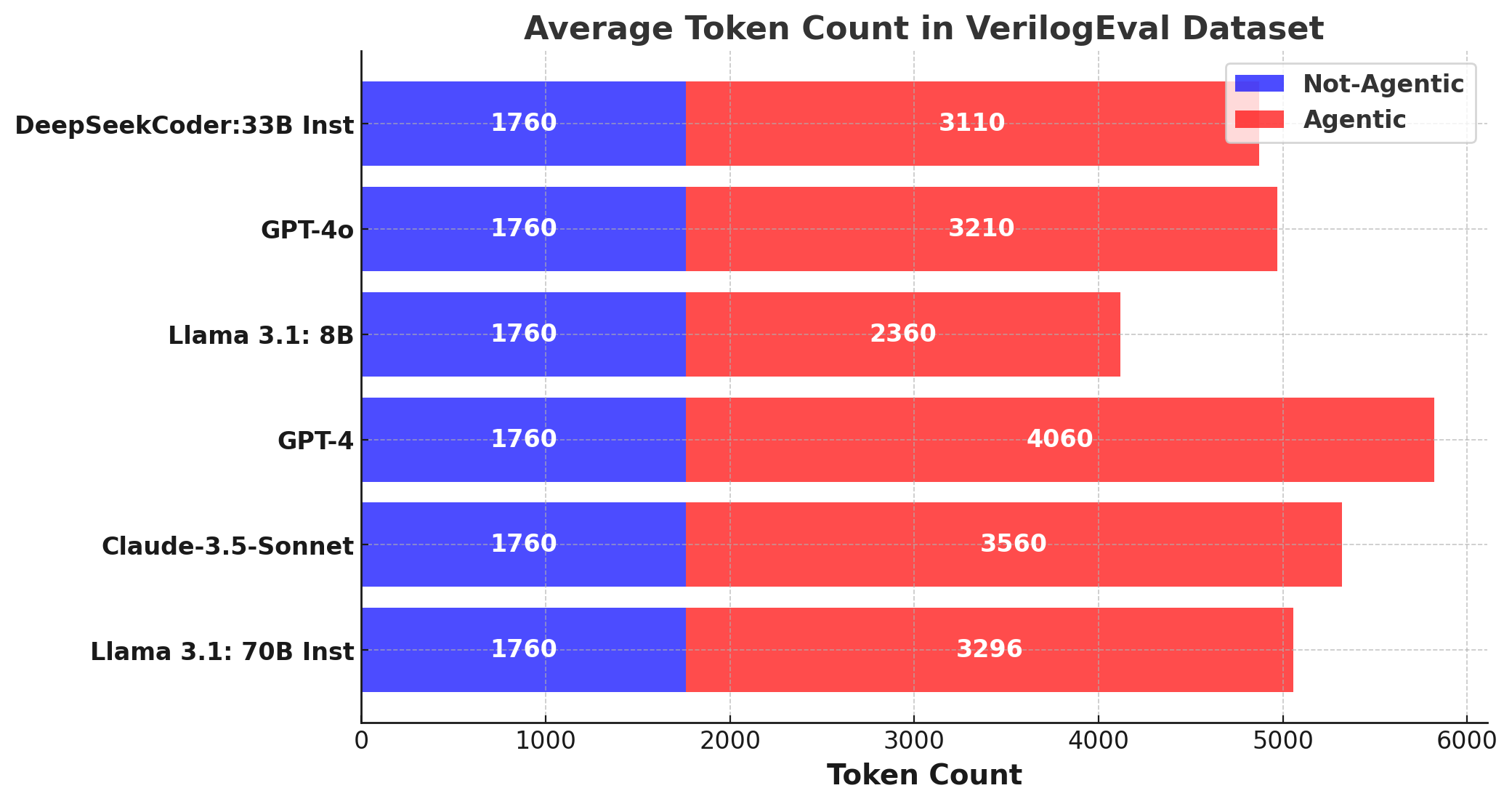}
    \caption{Average total token count across problems in the VerilogEval dataset(spec-to-rtl and code complete) after four LLM calls (excluding calls to the small Log\_summarizer agent)}
    \label{fig:tcbarplot}
\end{figure}

\subsection{Token Count comparison}
\label{sec:appendixtokencounts}

The bar plot shown in Figure \ref{fig:tcbarplot} presents the average total token count across problems in the VerilogEval dataset after four LLM calls, distinguishing between Not-Agentic and Agentic processing. Agentic models consistently use more tokens than their Not-Agentic counterparts due to increased reasoning, verbosity in the \textit{log\_inputs}, and more detailed outputs. \textit{Claude-3.5-Sonnet} (Agentic) exhibits the highest token usage (3560), followed by GPT-4 (4060) and GPT-4o (3210), suggesting extensive response expansion, such as additional explanations. Llama 3.1: 70B Inst (Agentic) and DeepSeekCoder:33B Inst (Agentic) show moderate token increases, whereas Llama 3.1: 8B (Agentic) has the lowest additional token usage (2360), indicating a more constrained agentic expression. While agentic models require a higher token count, this overhead contributes to improved problem-solving success and reduced LLM calls, ultimately leading to greater efficiency.

\end{document}